\documentclass[journal]{IEEEtran}
\usepackage{amsmath}
\usepackage{amssymb}
\usepackage{amsfonts}
\usepackage{algorithm}
\usepackage{graphicx}
\usepackage{epsfig}
\usepackage{subfigure}
\usepackage{psfrag}
\usepackage{xcolor}
\usepackage{cite}

\title{Massive MIMO for Cellular-Connected UAV: Challenges and Promising Solutions}
\author{\IEEEauthorblockN{Yi Huang, Qingqing Wu, Rui Lu, Xiaoming Peng, and Rui Zhang}
}

\begin{document}
\maketitle \thispagestyle{empty} \vspace{-0.3in}

%\begin{abstract}
%\end{abstract}
%
%
%\begin{IEEEkeywords}
%\end{IEEEkeywords}

\newtheorem{definition}{Definition}
\newtheorem{assumption}{Assumption}
\newtheorem{lemma}{\underline{Lemma}}
\newtheorem{example}{Example}
\newtheorem{theorem}{Theorem}
\newtheorem{proposition}{Proposition}
\newtheorem{conjecture}{Conjecture}
\newtheorem{remark}{Remark}
\newcommand{\mv}[1]{\mbox{\boldmath{$ #1 $}}}

%\section{Introduction}\label{sec:Introduction}

\begin{abstract}
Massive multiple-input multiple-output (MIMO) is a promising technology for enabling cellular-connected unmanned aerial vehicle (UAV) communications in the future. Equipped with full-dimensional large arrays, ground base stations (GBSs) can apply adaptive fine-grained three-dimensional (3D) beamforming to mitigate the strong interference between high-altitude UAVs and low-altitude terrestrial users, thus significantly enhancing the network spectral efficiency. However, the performance gain of massive MIMO critically depends on the accurate channel state information (CSI) of both UAVs and terrestrial users at the GBSs, which is practically difficult to achieve due to UAV-induced pilot contamination and UAV's high mobility in 3D. Moreover, the increasingly popular applications relying on a large group of coordinated UAVs or UAV swarm as well as the practical hybrid GBS beamforming architecture for massive MIMO further complicate the pilot contamination and channel/beam tracking problems. In this article, we provide an overview of the above challenging issues, propose new solutions to cope with them, and discuss about promising directions for future research. Preliminary simulation results are also provided to validate the effectiveness of proposed solutions.
\end{abstract}

\section{Introduction}\label{sec:Introduction}
In the past decade, there has been an increasing demand for unmanned aerial vehicles (UAVs) in a proliferation of new applications, such as aerial photography, cargo delivery, surveillance, airborne communication, etc. The basic communication requirements for UAVs can be classified into two types: the control and non-payload communication (CNPC) for command and control message that requires high reliability and low latency, and the payload communication for application data such as high-rate video streaming. As compared to the existing UAV-ground communications over the unlicensed spectrum (e.g., 2.4 GHz) that are unreliable, insecure, and of limited range and data rate, the almost ubiquitous connectivity and ultra-low-latency backhaul of the fifth-generation (5G) cellular network are promising to enable far more reliable and significantly higher capacity communication links for UAVs in a cost-effective manner \cite{ZengTutorial}.

Despite its great potential, cellular-connected UAV brings new challenges in operating 5G networks in the future, highlighted as follows.
\begin{itemize}
\item \textbf{Severe Air-Ground Interference:} The wireless channels between high-altitude UAVs and ground base stations (GBSs) are typically dominated by the strong line-of-sight (LoS) path \cite{ZengTutorial}. {As shown in Fig. 1(a), although} LoS links bring high degrees of macro-diversity for UAVs to potentially connect with more available GBSs, they inevitably  cause severe interference to the concurrent transmissions of the ground user equipments (GUEs) over the same frequency channels in the uplink, and also render the UAVs  more vulnerable to the interference from the co-channel GBSs transmitting to GUEs in the downlink.

\item \textbf{Ubiquitous Three-Dimensional (3D) Coverage for UAVs:} {As shown in Fig. 1(a), although} UAVs usually operate in much higher altitude than GUEs, thus requiring the GBSs to offer 3D signal coverage for them. However, the GBS antennas in the existing 4G network are typically tilted downward to provide ground coverage only while reducing the inter-cell interference among GUEs. This thus results in insufficient coverage for communications with UAVs in the sky.
\end{itemize}

\begin{figure}
\centering
\includegraphics[width=9cm]{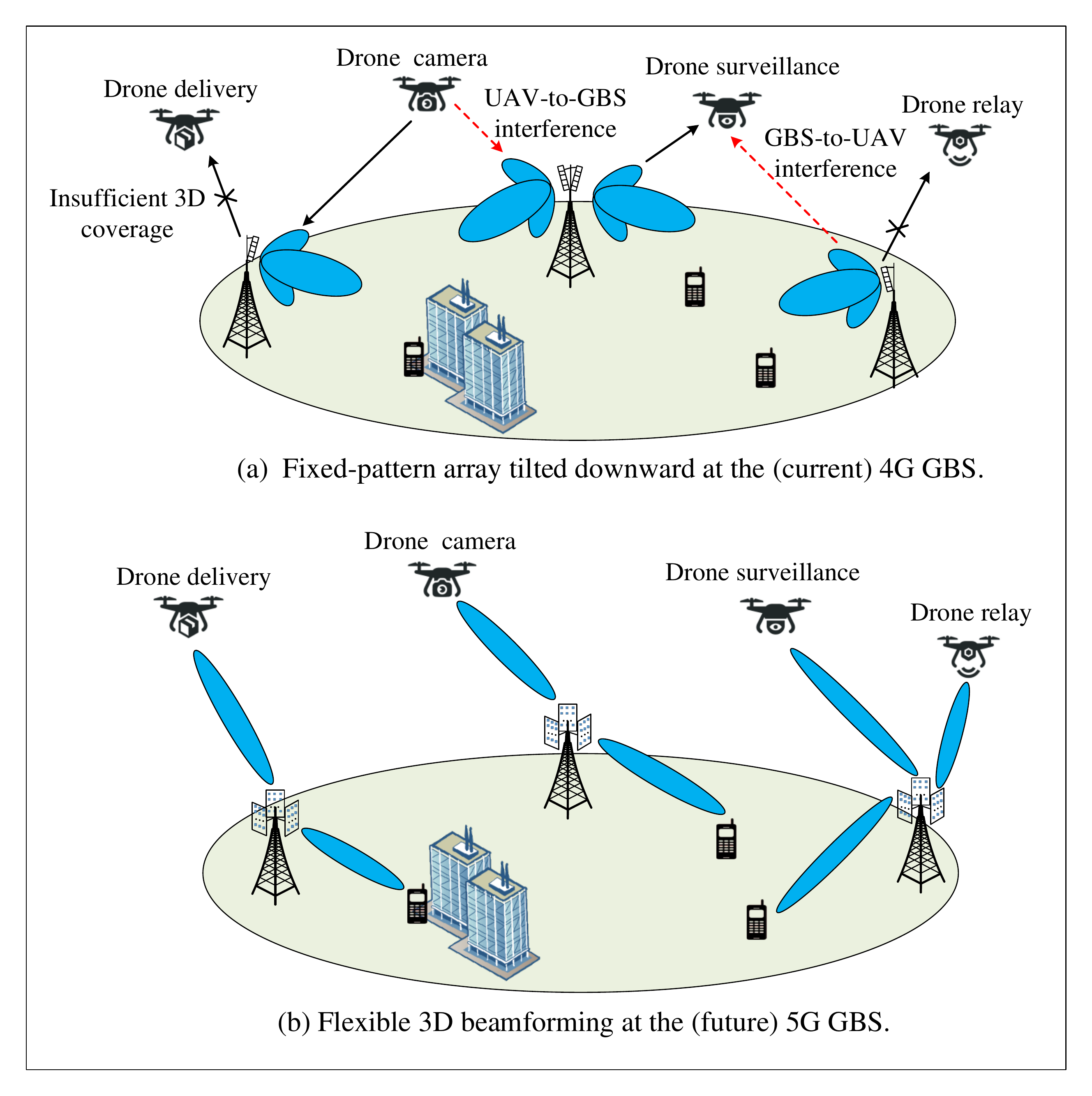}
\caption{4G/5G cellular networks serving both UAVs and GUEs.}
\label{fig1}\vspace{-5mm}
\end{figure}

Massive multiple-input multiple-output (MIMO) is a key technology that dramatically improves the spectral efficiency of 5G wireless systems, by leveraging  a large number of GBS antennas to serve a small number of users. Due to its enormous beamforming and full spatial multiplexing gains, massive MIMO has been applied recently  to support UAVs for resolving the aforementioned aerial-ground interference and 3D coverage issues \cite{Bjornson2,LarssonSwarm,UAVsensor}. As compared to traditional two-dimensional (2D) beamforming towards the ground only, massive MIMO with 3D beamforming provides finer-grained angle resolutions in both azimuth and elevation dimensions{. Thus, it} offers far more effective interference mitigation capability in both the UAV uplink and downlink communications by exploiting the elevation angle difference between UAVs and GUEs \cite{ZengTutorial}, {as shown in Fig. \ref{fig1}(b)}. Besides, 3D beamforming improves connectivity and coverage for UAVs in the sky due to more flexible elevation-domain beamforming.

{Although massive MIMO is promising for interference suppression and coverage extension in cellular-connected UAV communications, it faces several practical challenges in serving UAVs, e.g., pilot contamination incurred by UAVs with strong LoS channels and channel/beam tracking for UAVs with 3D high mobility. Moreover, serving the UAV swarm and applying practical hybrid GBS beamforming architecture for massive MIMO further complicate pilot contamination and channel/beam tracking problems. This thus motivates this article to provide an overview on the new issues and challenges in massive MIMO for supporting UAV communications. New and promising solutions are also proposed to cope with them. Moreover, numerical examples {are presented} to demonstrate the effectiveness of {the proposed} solutions.}

\section{Challenges for Serving UAVs with Massive MIMO}
{The performance of massive MIMO depends critically on the accuracy of CSI for all served users (UAVs and GUEs) at each GBS, which faces several new issues outlined as follows.}

{\subsection{\textbf{UAV-Induced Pilot Contamination}}}
Pilot contamination arising from multi-cell pilot reuse may cause severe interference regardless of the number of GBS antennas \cite{PilotDecontam1}. To address this issue, using a large pilot reuse factor is an effective solution for terrestrial cellular networks, since it ensures that the GUEs sharing the same pilot are sufficiently far apart, thus rendering their pilot contamination negligible. However, the pilot contamination problem becomes more challenging to solve in cellular-connected UAV communications. {Specifically,} UAV is usually at a much higher altitude than GBSs and thus has strong LoS channels with them in a much wider area as compared to GUE \cite{JR:hua20203d}. {Thus, UAV causes} severe pilot contamination with more users (GUEs as well as other UAVs) that reuse the same pilot, even when their serving GBSs are far from this UAV. Therefore, how to deal with the UAV-induced pilot contamination is a new challenge in massive MIMO.

{\subsection{\textbf{3D Beam Tracking for UAV}}}
In order to fully exploit the beamforming gain by massive MIMO and provide 3D coverage for a flying UAV, its serving GBS needs to obtain accurate CSI for it over time via efficient beam tracking. {Different from the terrestrial fading channels of GUEs with GBSs, UAVs have LoS links with their serving GBSs at most time via dynamic GBS association. Therefore, the serving GBSs for UAVs need to be able to simultaneously track the azimuth and elevation angles of their LoS links with the UAVs}. Moreover, the high and 3D mobility of UAVs causes rapid channel phase variations over time and thus incurs more frequent pilot transmissions for channel/beam tracking. {Note that the UAV-mounted Global Positioning System (GPS) is costly to implement in practice to facilitate beam tracking. Even if the GPS is available,  its precision (in the order of meters) is insufficient for accurate beam tracking and can only provide coarse location information for helping reduce the training overhead. As a result, how to achieve efficient and precise 3D beam tracking for UAVs by exploiting their unique channel characteristics with GBSs is a new and practically important problem.}

{\subsection{\textbf{Challenges in Serving UAV Swarm}}}
The UAV swarm formed by a large group of coordinated UAVs can accomplish more sophisticated  tasks than a single UAV and thus has attracted growing interest recently. However, obtaining accurate CSI for UAV swarm is also more challenging for the serving GBS. On one hand, a large number of UAVs in the UAV swarm generally require more pilots assigned to them, thus causing more severe pilot contamination and resulted interference with GUEs and other UAVs. On the other hand, tracking the channels of such a large number of UAVs also incurs more pilot overhead, which would decrease the communication throughput. To tackle the above problems, more efficient channel estimation and communication designs for UAV swarm are imperative.

{\subsection{\textbf{Considerations for Practical Hybrid Beamforming}}}
The so-called hybrid digital and analog beamforming based on the sub-array architecture has been widely adopted for implementing massive MIMO in 5G \cite{FDmimo}. To find the best analog beamforming directions towards users, 5G GBSs usually perform exhaustive beam searching with multiple pilot symbols sent by the users. However, if these beam searching pilots are contaminated by the interfering UAVs, the GBS will fail to find the right analog beamforming direction towards its served user, and thus beam misalignment occurs and low signal-to-noise ratio (SNR) is resulted. Thus, how to tackle the new challenges due to UAVs as well as the practical hybrid beamforming at GBSs is also crucial.

\
\section{Pilot Decontamination in Cellular-Connected UAV Communication}
\begin{figure}
\centering
\includegraphics[width=9cm]{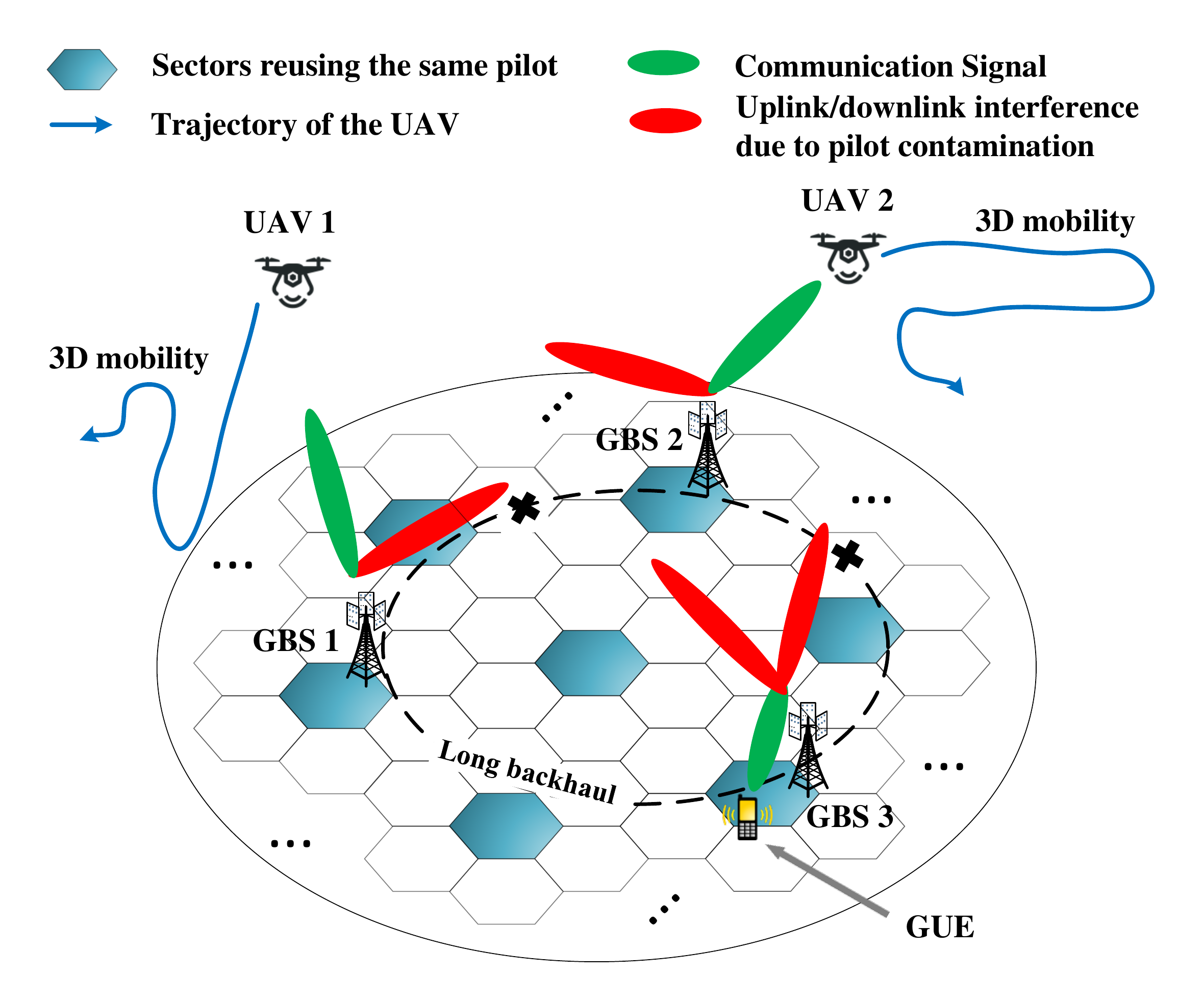}
\vspace{-8mm}
\caption{Pilot contamination in the cellular network serving both UAVs and GUEs.}
\label{fig2}\vspace{-4mm}
\end{figure}

Fig. \ref{fig2} shows the UAV-induced pilot contamination and resulted interference in a massive MIMO network due to the typically strong LoS channels between UAVs and GBSs in a large area. It is worth noting that traditional pilot decontamination techniques for GUEs, such as large-scale fading based precoding scheme \cite{PilotDecontam1} or coordinated pilot assignment scheme \cite{PilotDecontam2}, are insufficient or even infeasible for cellular-connected UAV communications, since in this case they will require cooperation between far apart GBSs and hence incur much more signaling overhead and longer backhaul delay {(see Fig. 2)}. Therefore, new and more efficient methods need to be devised to eliminate the effect of UAV-induced pilot contamination, as detailed in the following.

\subsection{Pilot Decontamination for GUE} When the number of UAVs in the network is not large, one effective method to alleviate their induced pilot contamination is by reserving a dedicated pool of pilots that are exclusively used by UAVs. In this case, GUEs will be free of pilot contamination caused by UAVs; however, this reserved pool of pilots may be under-utilized in an area with very few UAVs communicating with the GBSs. To tackle this problem, dynamic pilot assignment could be adopted, where the pilots unused by any UAVs are temporarily assigned to GUEs. In this way, the pilot efficiency can be improved without causing any pilot contamination between UAVs and GUEs, although a certain network-level coordination for pilot assignments is required.

Notwithstanding, in future cellular networks with both high-density UAVs and GUEs, all pilots in the network will need to be shared/reused among them. In this case, the pilot contamination between them can be resolved with spatial interference cancellation techniques implemented at each GBS independently, by exploiting their different channel characteristics with the GBS \cite{LuRui}. For instance, if the received pilot from a GUE is contaminated by one or more UAVs, its serving GBS can first detect the LoS directions of these interfering UAVs by applying a spatial matched filter over the possible azimuth and elevation angle ranges of UAVs, and then project the contaminated GUE's channel estimate into the null space of the detected UAVs' LoS directions to eliminate the pilot contamination. This is practically feasible and effective since channel subspaces of GUEs can be easily differentiated from those of UAVs, thanks to the sufficient separation of their elevation angles in 3D space.

\subsection{Pilot Decontamination for UAV} Pilot contamination among UAVs is more challenging than that between them and GUEs because they have similar (LoS) channel characteristics with GBSs. Note that the interference cancellation technique for resolving the UAV-induced pilot contamination with GUEs cannot be directly applied for that among UAVs. This is because the angle of arrival (AoA) of the desired UAV user's LoS channel with its serving GBS cannot be separated from those of the LoS channels from other non-associated UAVs that use the same pilot. To detect the LoS channels and extract their AoAs for all non-associated UAVs using the same pilot, each GBS can let its served UAV send one or more additional pilots with different training sequences if the pilot contamination is detected after the first pilot transmission (i.e., more than one dominant AoAs have been detected) \cite{LuRui}. The training sequences are selected randomly and independently over GBSs so that the other non-associated UAVs that have caused the pilot contamination during the first pilot transmission are unlikely to use all the subsequently sent training sequences for a given UAV. Thus, the unique LoS path of its served UAV can be detected as the only common LoS path present over all the pilots received at each GBS, at the cost of slightly more pilot overhead and higher computational complexity as compared to the case of GUEs. The number of additional pilots used should be chosen in practice to achieve a balance between pilot overhead and pilot decontamination performance.

There are still remaining issues for resolving UAV-induced pilot contamination in practice. For example, if GUEs are at higher altitude and/or UAVs fly at lower altitude, it will be more difficult to differentiate their channels by exploiting different elevation-domain AoAs. {Moreover, the  non-LoS (NLoS) components in UAV-GBS channels may affect the performance of the proposed pilot-decontamination schemes and its impact is worthy of further investigation.}

\section{3D Beam Tracking for Cellular-Connected UAV}
To provide reliable coverage for the served UAVs as well as mitigate the interference from other non-associated UAVs after initial pilot decontamination (as discussed in the preceding section), each GBS needs to accurately track the channel/beam directions of its served UAVs over time. Conventionally, this can be achieved by continuously estimating the channels of UAVs based on the periodic pilots sent by them. However, since UAVs usually have high mobility in 3D space (see Fig. 2), their channels with associated GBSs suffer rapid phase variations over time, which requires frequent pilot transmissions. Therefore, more efficient 3D beam tracking schemes need to be designed for UAVs in the massive MIMO network to reduce the pilot overhead and enhance the communication throughput by exploiting their LoS channels with associated GBSs, as will be addressed in this section. {Note that even if the UAV's LoS path with its associated GBS may be occasionally blocked, it can still be connected to another GBS with an LoS link by exploiting the macro diversity in the cellular network.}

\subsection{{Channel Prediction Based on Angular Speeds Estimation}}
In practice, each GBS can acquire the knowledge of its served UAV's velocity and thereby predict the LoS channel between them over time to save pilot transmissions. To this end, {one may} assume that the UAV's angular speeds, i.e., the instantaneous variation rates of its azimuth and elevation angles, stay approximately unchanged during a given short period \cite{FirstOrder}. As such, each GBS can first detect the azimuth and elevation angles of its served UAV's LoS path over two non-consecutive beam tracking pilots received from the UAV, and then estimate the UAV's angular speeds as the variations of its azimuth and elevation angles normalized to the time difference between the two pilot symbols, respectively. Based on the estimated angular speeds, the UAV's azimuth/elevation angles as well as its LoS channel in subsequent time slots can be predicted without the need of sending new pilots. The above algorithm can be implemented with low complexity yet without the help of conventional positioning technologies such as GPS. However, the LoS channel prediction becomes inaccurate once the UAV's angular speeds change. As a result, the beamforming gain at the GBS degrades over time. To overcome this issue, the GBS can estimate the angular speeds of the UAV from time to time based on its periodically sent pilots.
\vspace{-5mm}
\subsection{{Kalman Filter Based Channel Tracking}}
Kalman filter based channel tracking\cite{Gao} is another effective method, where the GBS iteratively performs channel prediction for its served UAV and updates the predicted channel based on independent channel measurement (e.g., that obtained from conventional channel estimation pilots). Note that although the algorithm presented in the previous subsection can be applied to the above channel prediction step, more sophisticated channel prediction algorithms can also be considered to improve the performance. For example, each GBS can estimate the UAV's velocity and acceleration jointly for channel prediction by modeling its mobility as a second-order dynamic system.

In general, the Kalman filter based method is able to obtain more accurate prediction of the UAV's channel as compared to the previously introduced method based on angular speeds estimation, since the channel prediction error in former case can be further reduced by exploiting the additional channel measurement in the update step. However, this requires more pilot overhead as well as higher processing complexity at the GBS. Besides, if UAV's velocity changes too rapidly, the channel prediction error will grow substantially even with the Kalman filter based channel tracking.

To further improve the accuracy and robustness of channel tracking for UAVs, multi-GBS cooperation is one  promising direction for future research.  Specifically, the LoS-dominant channels of a UAV with multiple GBSs bring a higher macro-diversity gain as compared to that of a GUE. Thus, the pilots sent by a UAV are likely to be received by neighbouring GBSs of its serving GBS over LoS channels, each of which can help estimate the UAV's velocity. As such, the UAV's serving GBS can collect  information on its velocity from these neighbouring GBSs to improve the channel prediction performance.

\section{Massive MIMO Communication with UAV Swarm}
\begin{figure}
\centering
\includegraphics[width=9cm]{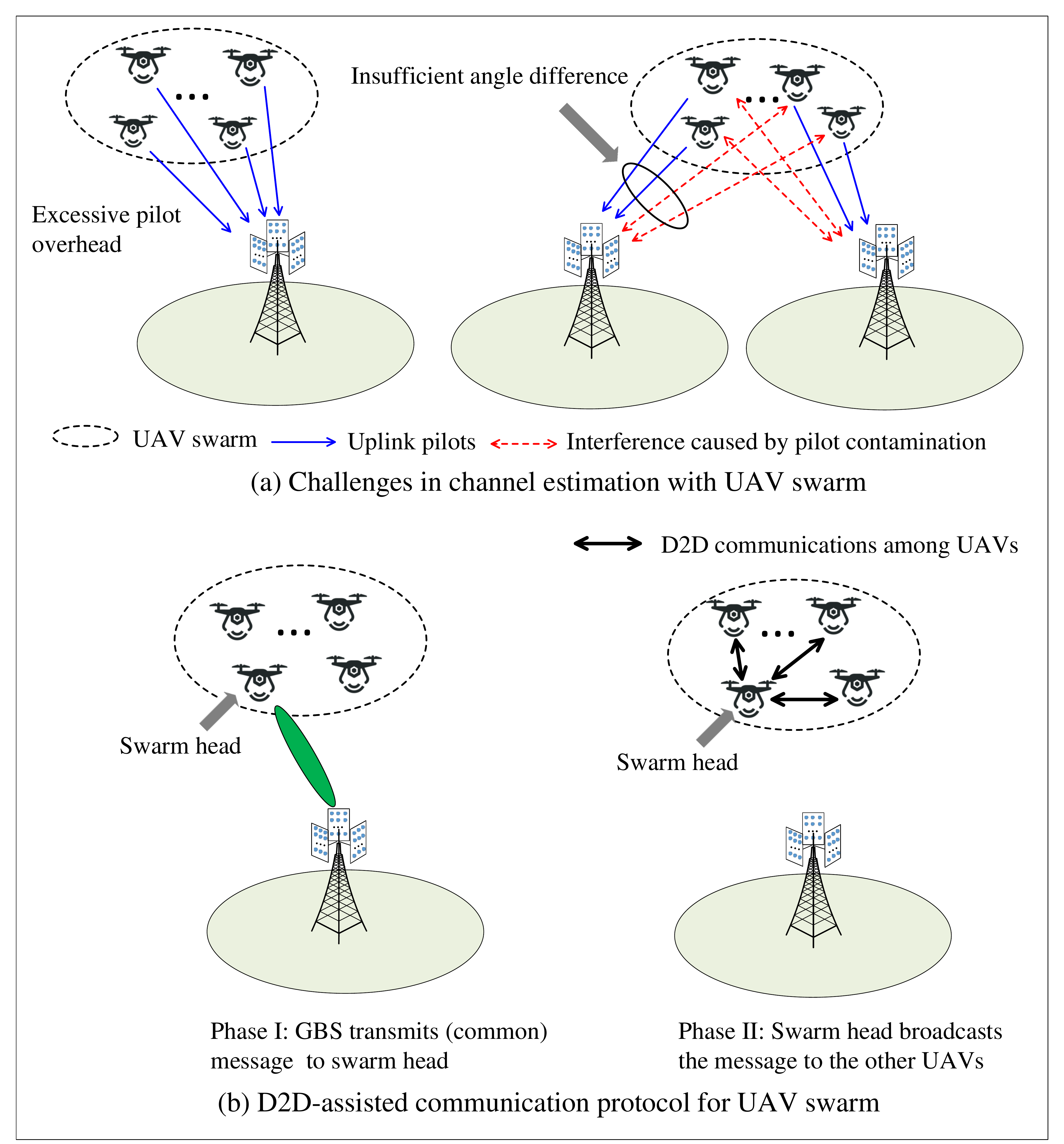}
\caption{{D2D based solution for supporting UAV swarm with massive MIMO.} }
\label{fig3}\vspace{-5mm}
\end{figure}

Although it is practically appealing to employ massive MIMO to support communications between GBS and UAV swarm \cite{LarssonSwarm}, the large number of UAVs in a swarm may require excessive pilots for channel estimation and thus cause more severe pilot contamination in the network, {as shown in Fig. \ref{fig3}(a)}. Moreover, the angle difference among UAVs in the same swarm is practically small. Thus, their LoS channels with the same serving GBS are more difficult to be differentiated as compared to the single-UAV case, which renders {the proposed} pilot decontamination methods in Section III less effective. Therefore, new and more effective channel estimation and communication designs are needed for supporting UAV swarm with massive MIMO.

One potential solution is to choose a UAV head in the swarm for channel estimation with the serving GBS only by sending a pilot in the uplink. In this way, the pilot overhead is substantially reduced, while the UAV-induced pilot contamination becomes the same as that in the single-UAV case and thus can be resolved by the methods proposed in Section III. After channel estimation, a two-phase transmission protocol can be adopted for enabling communications between the GBS and its associated UAV swarm, {as shown in Fig. \ref{fig3}(b)} for the case of downlink communication. In phase I, the GBS beams toward the swarm head based on the (uplink) estimated CSI for sending a (common) message to it. In phase II, device-to-device (D2D) communications among the UAVs are utilized such that the swarm head broadcasts its received message to the other UAVs for them to retrieve their intended information from the common message. Similarly, D2D-assisted uplink communication from UAV swarm to GBS can be implemented.

There are interesting aspects of the above D2D-assisted massive MIMO communication with UAV swarm worthy of further investigation. Firstly, the durations of the two phases subject to a given transmission delay need to be properly set to strike an optimal balance between the achievable rates in the two phases so as to maximize the system throughput. Secondly, when the number of UAVs in the swarm is excessively large, multiple swarm heads may be appointed for reducing the D2D transmission delay in phase II. However, this may incur over-the-air interference among the swarm heads as well as more severe pilot contamination in phase I. Thus, optimal design of swarm heads and their D2D transmission protocol is also crucial to maximize the effective system throughput. {For example, a UAV head can be selected based on higher probability of LoS path with the GBS for improving the link quality. Furthermore, it should consume minimum power to communicate with the rest of swarm members over the wireless links between them \cite{Backhaul}.}

\section{Practical Considerations for Hybrid Beamforming}
Low-cost hybrid analog and digital beamforming with much less RF chains required than full-digital beamforming is commonly adopted in massive MIMO systems. There are two practical architectures for implementing it at the GBS, namely, full-complexity or reduced-complexity \cite{FDmimo}, {as shown in Figs. 4(a) and 4(b) respectively}. In the full-complexity case, each RF chain is connected to all antenna elements and thus the GBS can combine multiple analog beams in the digital domain (where each analog beam corresponds to an RF chain). While in the reduced-complexity case, each RF chain is exclusively connected to a sub-array and all sub-arrays perform analog beamforming independently.

For both full-complexity and reduced-complexity architectures, the beam searching pilots received by the GBS may be contaminated by non-associated UAVs with strong LoS channels. Moreover, the interference cancellation based pilot decontamination proposed in Section III becomes ineffective since the number of RF chains at the GBS is insufficient to detect the LoS directions of the interfering UAVs directly. Thus, new solutions are needed to resolve pilot contamination between UAVs and GUEs as well as among UAVs in beam searching. One possible solution is that the GBS can restrict the angle-searching range for its associated GUE towards the ground only, such that the pilot contamination between the GUE and non-associated UAVs can be effectively alleviated. Meanwhile, when the pilot of the associated UAV is contaminated by non-associated UAVs, the GBS can let its associated UAV send additional pilots randomly selected for beam searching, and thereby the desired beam direction of its associated UAV {can be found} as the common one detected from all the pilots received (similarly as in Section III for differentiating the LoS path of the associated UAV from those of non-associated UAVs). Alternatively, the GBS can send downlink beam searching pilots in the directions found in the uplink training, and its associated UAV then sends the index of the desired beam direction back to the GBS, which only requires few additional pilots in practice.
\begin{figure}
\centering
\includegraphics[width=8cm]{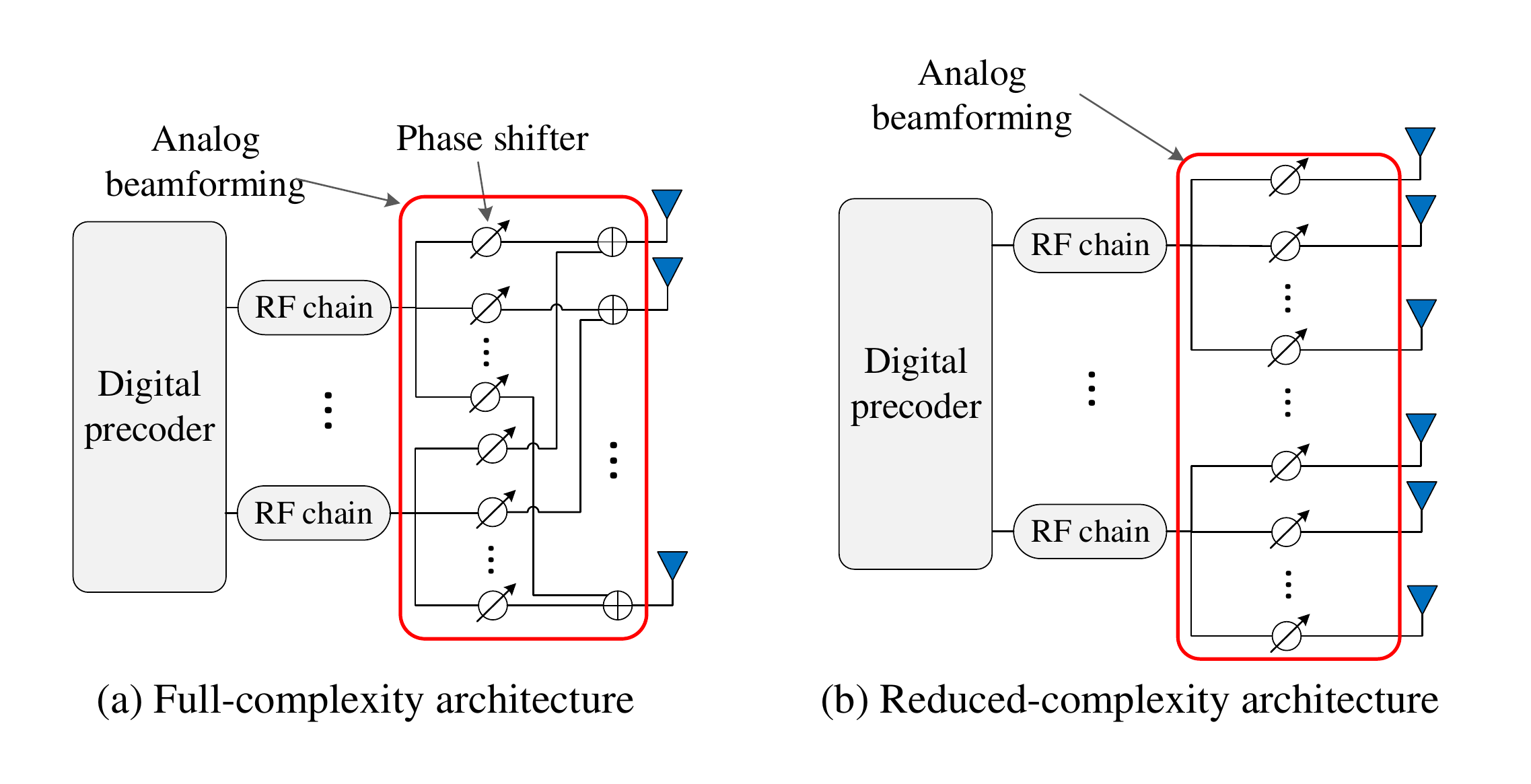}
\vspace{-3mm}
\caption{Illustration of {practical hybrid beamforming architecture}.}
\label{fig4}\vspace{-5mm}
\end{figure}

On the other hand, due to 3D mobility of UAVs,  beam searching in both azimuth and elevation dimensions are generally required, which incurs more pilot overhead as compared to GUEs. {For reducing pilot overhead in beam searching, one practical approach is to use the hierarchical multi-resolution codebook \cite{Xia, Yanglu}.} Starting with searching over the low-resolution beam directions in the codebook, the GBS first obtains the coarse beam direction in azimuth/elevation dimension for its served UAV, then refines it using the high-resolution beam directions in the codebook. {However, the UAV with 3D and high mobility may cause fast channel variation and hence the optimal beam varies quickly, which still requires frequent beam searching. To tackle this problem, analog beam prediction based on Kalman filter for millimeter-wave communications \cite{Love} can be applied for tracking the UAVs.}

There are still remaining issues in communicating with UAVs via hybrid beamforming in massive MIMO, which is worth further investigating in the future. For example, how to solve the beam searching pilot contamination issue and improve beam searching efficiency for UAV swarms is still an open problem. Besides, joint optimization for channel estimation and data transmission for UAV swarms with hybrid beamforming is crucial but unaddressed yet.

\section{Simulation Results and Discussions}
In this section, simulation results {are presented} to verify the effectiveness of the pilot decontamination and 3D beam tracking techniques proposed in Sections III and IV, respectively. Consider a cellular system consisting of $7$ macro sites with $3$ sectors per site. Moreover, {it is assumed} that each sector is equipped with an $8\times16$ uniform planar array (UPA) with $8$ and $16$ antennas in vertical and horizontal dimensions, respectively. {Full-digital beamforming is assumed for evaluating the performance of proposed pilot decontamination and 3D beam tracking methods.} The carrier frequency is {$3.5\,$GHz} and the channel bandwidth is {$1\,$MHz}. The altitude of each GUE is fixed to {$1.5\,$meter} (m), whereas that of each UAV is uniformly distributed between {$50\,$m and $300\,$m}. The GBS's transmit power is {$46\,$dBm} while the user's transmit power is {$23\,$dBm}. The noise power density is {$-174\,$dBm/Hz} with additional noise figure of {$9\,$dB} in the downlink and {$5\,$dB} in the uplink. The 3GPP urban macro (UMa) scenario is considered with the same channel modeling as adopted in \cite{Zeng}.

\subsection{Pilot Decontamination}
{Consider} a given frequency channel reused by 21 users (15 UAVs and 6 GUEs) uniformly distributed over the sectors for evaluating {the proposed} pilot decontamination schemes. {The users are associated to the GBSs based on maximum reference signal receiving power (RSRP) principle with maximum ratio transmission (MRT) beamforming as described in \cite{Zeng}.} {It is assumed} that the users send pilot symbols in the uplink for channel estimation at their respective GBSs and the estimated uplink channels are treated as their corresponding downlink channels (i.e., assuming the uplink-downlink channel reciprocity as in typical time-division duplexing (TDD) based massive MIMO). {In the simulation, Zadoff-Chu sequences with length 12 are used for the pilots, which are generated by adding different cyclic shifts on a root sequence.} The pilot reuse factor is set to be $7$, i.e., the total number of orthogonal pilot sequences is $7$ and each sequence is reused by $3$ sectors. Note that with such a large pilot reuse factor, the pilot contamination among GUEs is negligible and thus can be ignored.

For comparison, the following three cases {are considered}: 1) with ideal CSI at GBSs for their associated users (i.e., without any pilot contamination among UAVs and GUEs), 2) with pilot contamination, and 3) with the interference cancellation based pilot decontamination as described in Section III applied for both GUEs and UAVs. {For detecting LoS paths of interfering UAVs, each GBS applies a spatial matched filter generated similarly as that in \cite{LuRui} over angle ranges $[0^\circ, 360^\circ]$ and $[0^\circ, 90^\circ]$ in azimuth and elevation dimension, respectively.}

\begin{figure}[t]
\centering
\includegraphics[width=8cm]{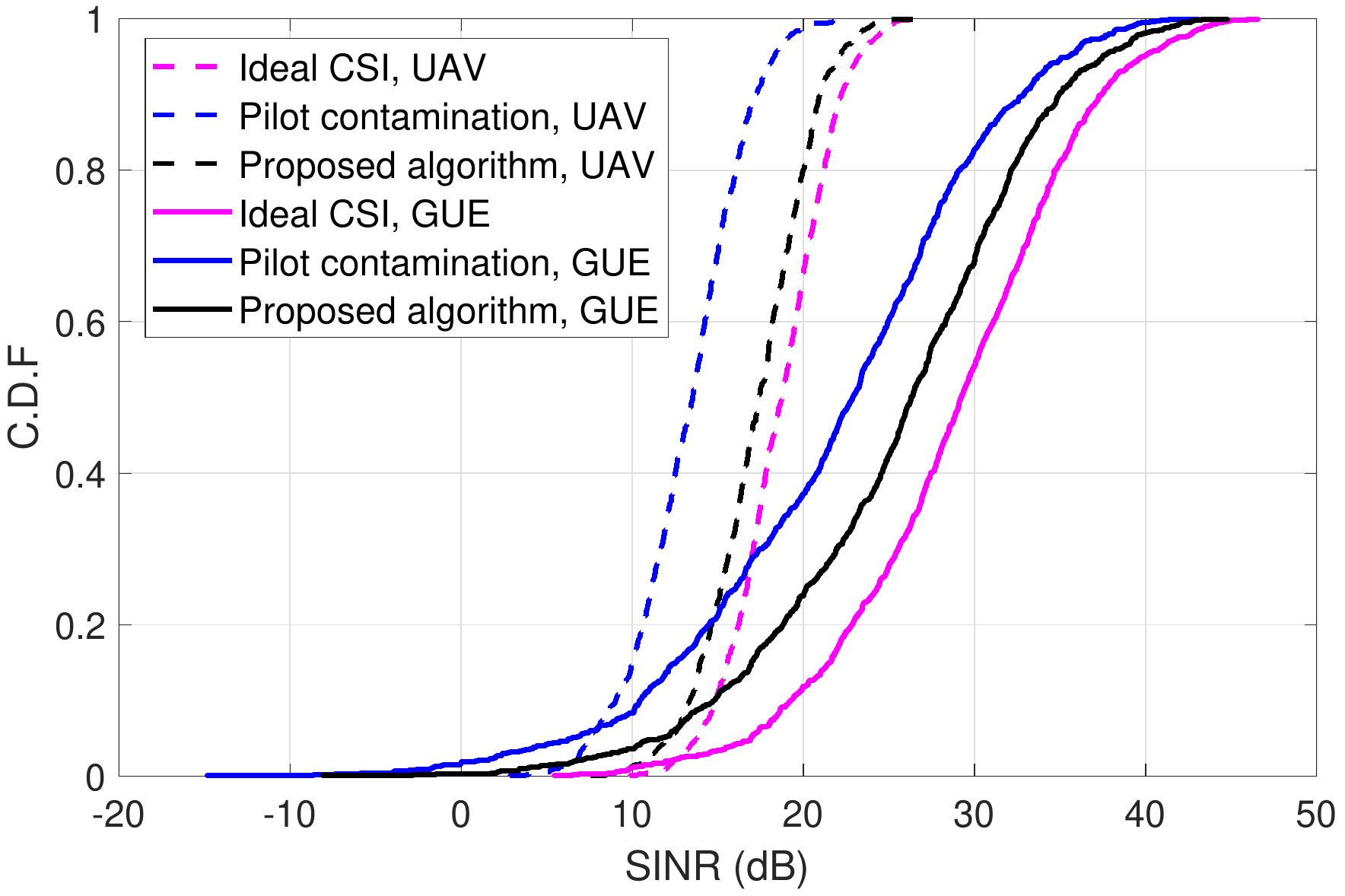}
\caption{SINR distribution of UAVs/GUEs in the downlink with/without pilot contamination versus that with the proposed pilot decontamination.}
\label{fig5}
\end{figure}

{Fig. \ref{fig5} shows} the downlink signal-to-interference-plus-noise ratio (SINR) distribution (in terms of cumulative distribution function (C.D.F.)) of different cases for the GUEs and UAVs, respectively. It is observed that the SINR of GUEs achieved by the proposed pilot decontamination is enhanced as compared to the case with pilot contamination. For example, the $5$th percentile SINR gain is {$5.4\,$dB} when the number of UAVs is $15$. This is because the power leakage to UAVs due to pilot contamination is alleviated effectively by the proposed scheme, and thus more signal power can be directed towards GUEs in the downlink transmission. In addition, it is also observed that the proposed pilot decontamination scheme for UAVs can improve their SINRs more significantly as compared to the case of GUEs. This is because the downlink interference due to pilot contamination impairs the UAV's SINR severely, while it has been substantially suppressed after effective pilot decontamination.

\subsection{3D Beam Tracking}
Next, {simulation results are provided to evaluate} the performance of 3D beam tracking algorithms for UAVs. {Consider} a GBS communicating with a single UAV in a time duration of {$4\,$seconds} (s). It is assumed that the UAV adopts a quasi-static mobility model and its velocity (including its moving direction and speed) changes randomly every {$1\,$s} \cite{FirstOrder}. The UAV's maximum speed is assumed to be {$160\,$kilometers per hour} (km/h) while the minimum speed is assumed to be {$40\,$km/h}.

\begin{figure}[t]
\centering
\includegraphics[width=8cm]{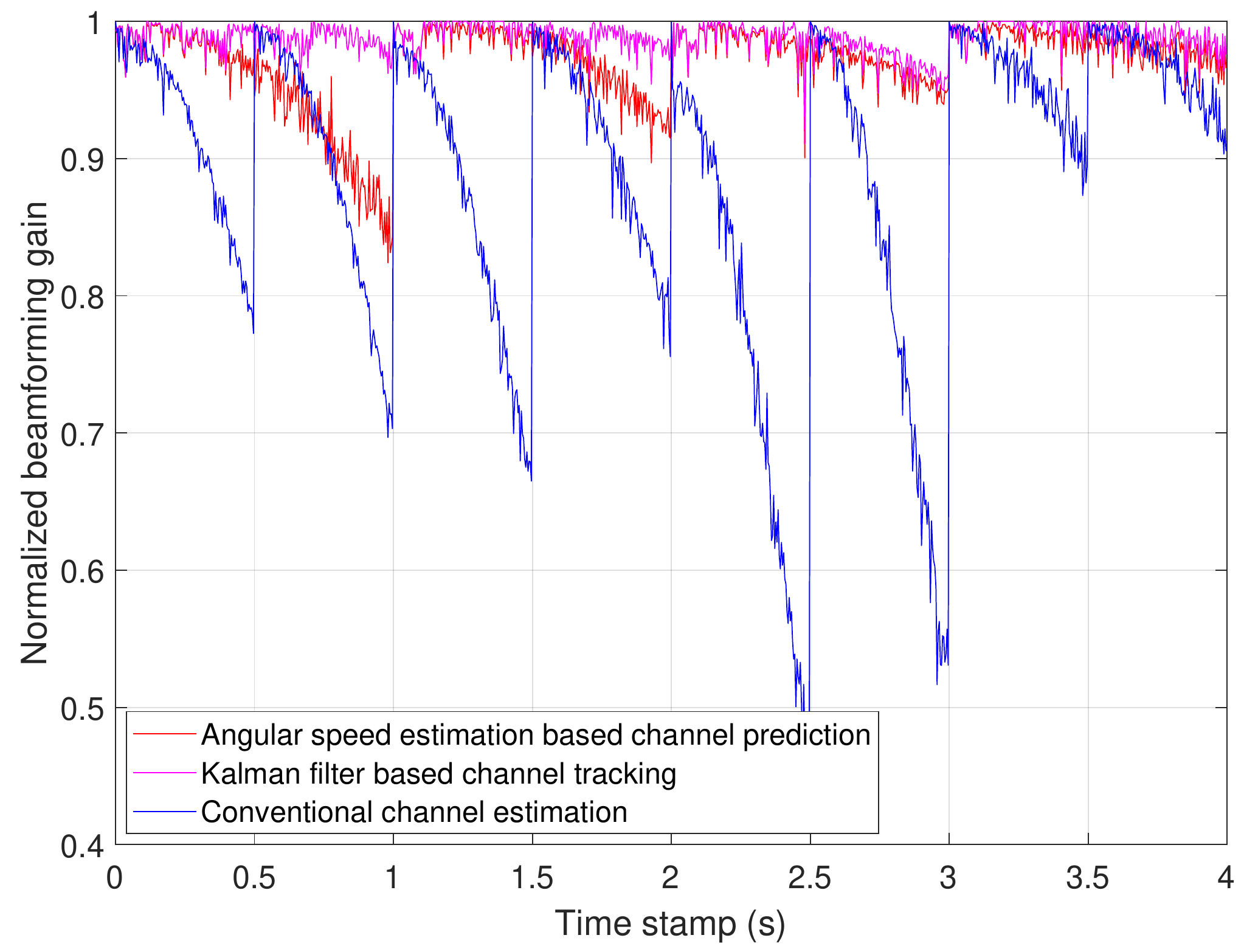}
\vspace{-1mm}
\caption{Performance comparison of 3D beam tracking algorithms.} \label{fig6}
\end{figure}

{Fig. \ref{fig6} shows} how the normalized beamforming gain (i.e., the beamforming gain under estimated CSI normalized by that under ideal CSI) changes over time under three schemes: 1) conventional channel estimation based on pilots sent by the UAV with a period of {$500\,$milliseconds} (ms); 2) channel prediction based on angular speeds estimation; 3) Kalman filter based channel tracking. {In Scheme 2, the UAV's angular speeds are estimated periodically based on the variations of its AoA between two pilot symbols with time interval {$100\,$ms}. In each pilot symbol, the AoA is detected at the GBS by applying a spatial matched filter similarly as that in pilot decontamination.} The period for angular speeds estimation is {$1\,$s}. {In Scheme 3, the GBS first performs channel prediction based on Scheme 2. Then the GBS updates the predicted channel by combining it with an independent channel measurement (which is obtained from additional uplink pilots with a period of {$500\,$ms}) based on Kalman filter.}

It is observed from Fig. \ref{fig6} that while the beamforming gain achieved by conventional channel estimation decreases fast over time, the beamforming gain can remain approximately unchanged by Scheme 2 (albeit the pilot overhead is the same for Schemes 1 and 2). Besides, since the UAV's angular speeds may change after each period of time which leads to inaccurate channel prediction, the beamforming gain achieved by Scheme 2 decreases occasionally but can be quickly recovered once the UAV's angular speeds are estimated again. Meanwhile, it is observed that Scheme 3 based on Kalman filter helps improve channel tracking accuracy as compared to simply performing channel prediction in Scheme 2, at the cost of more pilot overhead and higher implementation complexity.

\section{Conclusions}
{This article provided} an overview of the major challenges arising in massive MIMO network serving UAVs in addition to conventional GUEs. By exploiting the LoS-dominated UAV-GBS channels, new and effective pilot decontamination and 3D beam tracking solutions {were proposed} to obtain accurate CSI of UAVs, which is crucial to achieve the full massive MIMO gains. Moreover, we {considered} more challenging scenario of serving UAV swarm with massive MIMO and {proposed} new designs based on D2D communication for improving the system performance. Promising solutions for CSI acquisition under practical hybrid beamforming architecture {were also presented}. {Besides, this article also pointed out promising directions for future research, e.g., multi-GBS cooperative beam tracking for UAVs, D2D communication design for UAV swarm, and how to serve UAV swarm under practical hybrid beamforming, etc.}  It is hoped that this article will provide useful guidance to design and implement future massive MIMO networks serving both aerial and terrestrial users efficiently.

\bibliographystyle{IEEEtran}

%\begin{IEEEbiographynophoto}{Yi Huang} is a Research Fellow with the ECE Department, National University of Singapore.\end{IEEEbiographynophoto}
%
%\begin{IEEEbiographynophoto}{Qingqing Wu} is an Assistant Professor with the University of Macau, China. \end{IEEEbiographynophoto}
%
%\begin{IEEEbiographynophoto}{Rui Lu} is a visiting Ph.D. student with the ECE Department, National University of Singapore. \end{IEEEbiographynophoto}
%
%\begin{IEEEbiographynophoto}{Xiaoming Peng} is with the Satellite, Aviation \& Maritime Division, Institute for Infocomm Research, A*STAR, Singapore. \end{IEEEbiographynophoto}
%
%\begin{IEEEbiographynophoto}{Rui Zhang} (F'17) is a Full Professor with the ECE Department, National University of Singapore. \end{IEEEbiographynophoto}

\end{document}